\documentclass[aps,twocolumn,epsf,floatfix,pre]{revtex4-1}
\usepackage{graphics,graphicx,epsfig}
\usepackage{amssymb,color}
\usepackage{epsf,epstopdf,wrapfig}
\usepackage{amsmath}
\usepackage{multirow}

\begin{document}

\title{Collective behaviour without collective order in wild swarms of midges}

\author{Alessandro Attanasi$^{*,\ddagger}$, Andrea Cavagna$^{*,\ddagger,\natural}$, Lorenzo Del Castello $^{*,\ddagger}$, Irene Giardina$^{*,\ddagger,\natural}$,  Stefania Melillo$^{*,\ddagger}$, Leonardo Parisi$^{*,\S}$, Oliver Pohl$^{*,\ddagger}$, Bruno Rossaro$^\flat$, Edward Shen$^{*,\ddagger}$, Edmondo Silvestri$^{*,\dagger}$, Massimiliano Viale$^{*,\ddagger}$}

\affiliation{$^*$ Istituto Sistemi Complessi, Consiglio Nazionale delle Ricerche, UOS Sapienza, 00185 Rome, Italy}
\affiliation{$^\ddagger$ Dipartimento di Fisica, Universit\`a\ Sapienza, 00185 Rome, Italy}
\affiliation{$^\natural$ Initiative for the Theoretical Sciences, The Graduate Center, City University of New York, 365 Fifth Avenue, 10016 New York, USA}
\affiliation{$^\S$ Dipartimento di Informatica, Universit\`a\ Sapienza, 00198 Rome, Italy}
\affiliation{$^\dagger$ Dipartimento di Fisica, Universit\`a\ di Roma 3, 00146 Rome, Italy}
\affiliation{$^\flat$ Dipartimento di Scienze per gli Alimenti la Nutrizione e l'Ambiente, Universit\`a\ degli Studi di Milano, 20133 Milano, Italy}

\begin{abstract}
 
Collective behaviour is a widespread phenomenon in biology, cutting through a huge span of scales, from cell colonies up to bird flocks and fish schools. The most prominent trait of collective behaviour is the emergence of global order: individuals synchronize their states, giving the stunning impression that the group behaves as one. In many biological systems, though, it is unclear whether global order is present. A paradigmatic case is that of insect swarms, whose erratic movements seem to suggest that group formation is a mere epiphenomenon of the independent interaction of each individual with an external landmark. In these cases, whether or not the group behaves truly collectively is debated.
Here, we experimentally study swarms of midges in the field and measure how much the change of direction of one midge affects that of other individuals. We discover that, despite the lack of collective order, swarms display very strong correlations, totally incompatible with models of noninteracting particles. We find that correlation increases sharply with the swarm's density, indicating that the interaction between midges is based on a metric perception mechanism. By means of numerical simulations we demonstrate that such growing correlation is typical of a system close to an ordering transition. Our findings suggest that correlation, rather than order, is the true hallmark of collective behaviour in biological systems.
\vspace{0.5 cm} 

 {\bf Author Summary:} {\it Our perception of collective behaviour in biological systems is
closely associated to the emergence of order on a group scale. For example, birds within a
flock align their directions of motion, giving the stunning
impression that the group is just one organism. Large swarms of midges,
mosquitoes and flies, however, look quite chaotic and do not exhibit any
group ordering. It is therefore unclear whether these systems are true
instances of collective behaviour. Here we perform the three
dimensional tracking of large swarms of midges in the field and find
that swarms display strong collective behaviour despite the absence of
collective order. In fact, we discover that the capability of swarms
to collectively respond to perturbations is surprisingly large,
comparable to that of highly ordered groups of vertebrates.}

\end{abstract}

\maketitle

\section*{Introduction}

Intuition tells us that a system displays collective behaviour when all individuals spontaneously do the same thing, whatever this thing may be. We surely detect collective behaviour when all birds in a flock fly in the same direction and turn at the same time \cite{krause+ruxton_02}, as well as when all spins in a magnet align, giving rise to a macroscopic magnetization \cite{cardy,parisi}. On the other hand, we would not say that there is any collective behaviour going on in a gas, despite the large number of molecules. The concept of collective behaviour seems therefore closely linked to that of emergent collective order, or synchronization. Indeed, explaining how order spontaneously arises from local inter-individual interactions has been one of the major issues in the field  \cite{okubo_86,camazine+al_01,sumpter_10}. 

The case of insect swarms is tricky in this respect. Several species in the vast taxonomic order Diptera (flies, mosquitoes, midges) form big swarms consisting largely of males, whose purpose is to attract females \cite{downes_69,sullivan_81}. Swarming therefore has a key reproductive function and, in some cases, relevant health implications, the obvious, but not unique, example being that of the malaria mosquito, {\it  Anopheles gambiae} \cite{nielsen_60,charlwood_80,manoukis_09}. It is well-known that swarms form in proximity of some visual marker, like a water puddle, or a street lamp  \cite{downes_69}. Swarming insects seem to fly independently around the marker, without paying much attention to each other (see Video S1).
For this reason, the question of whether swarms behave as truly collective systems is debated \cite{okubo_86,okubo_01}. In fact, it has even been suggested that in Diptera there is no interaction between individuals within the swarm and therefore no collective behaviour at all \cite{downes_55,blackwell+al_92}.  Although other studies observed local coordination between nearest neighbors \cite{okubo_74,butail+al_13}, it remains controversial whether and to what extent collective patterns emerge over the scale of the whole group. Clarifying this issue is a central goal in swarms containment \cite{ali_80,rose_01}. In absence of quantitative evidence telling the contrary, the hypothesis that external factors, as the marker, are the sole cause of swarming and that no genuine collective behaviour is present, is by far the simplest explanation.

\begin{figure*}
\includegraphics[width=0.9\textwidth]{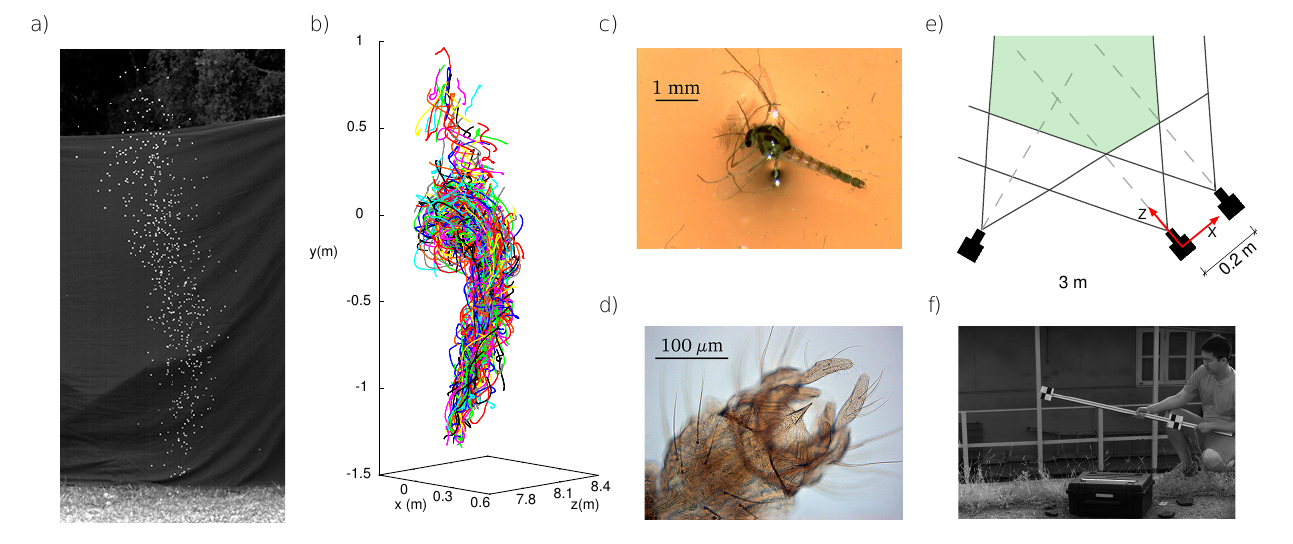}
\caption{
{\bf Experiment}.
{\bf a:} A natural swarm of midges ({\it Cladotanytarsus atridorsum}, Diptera:Chironomidae), in Villa Ada, Rome. The digital image of each midges is, on average, a $3\times 3$ pixels
light object against a dark background. 
{\bf b:} The $3d$ trajectories reconstructed 
for the same swarm  as in {\bf a}. Individual trajectories are visualized
for a short time, to avoid visual overcrowding (see also SM-Video 1 and 2).
{\bf c:} A microscope image of an adult male of {\it Cladotanytarsus atridorsum}. {\bf d:} A detailed view of the hypopygium, used for species identification
(see Methods); the same midge as in {\bf c}.
{\bf e:} Scheme of the experimental set-up. Three synchronized cameras recording at 
$170$ frames per second are used. Two cameras $3$ m apart are used as the stereoscopic pair  
for the three dimensional reconstruction. The third one is 
used to reduce tracking ambiguities and resolve optical occlusions. Three dimensional trajectories 
are reconstructed in the reference frame of the right stereoscopic camera. 
{\bf f:} The mutual geometric positions and orientations of the cameras are retrieved by taking 
several pictures of a known target. The accuracy we achieve in the determination of the mutual camera 
orientation is of the order of $10^{-4}$ radians.
}
\label{fig:experiment}
\end{figure*}

Physics, however, warns us to be careful in identifying collective behaviour with collective order. There are systems displaying important collective effects both in their ordered {\it and} in their disordered phase. The most important example is that of a ferromagnet: the collective response to an external perturbation is as strong in the disordered phase slightly above the critical temperature $T_c$, i.e. the temperature below which a spontaneous magnetization emerges,  as it is in the ordered phase slightly below $T_c$. In fact, once below the critical temperature, increasing the amount of order {\it lowers} the collective response \cite{cardy,parisi}. Hence, although certainly one of its most visible manifestations, emergent order is not necessarily the best probe of collective behaviour. 

The crucial task for living groups is not simply to achieve an ordered state, but to respond collectively to the environmental stimuli. For this to happen, correlation must be large, namely individuals must be able to influence each other's behavioural changes on a group scale. The question then arises of whether correlation in biological systems is a consequence of collective order or whether it can be sustained even in absence of order, as it happens in some physical systems. 
The question is relevant because the way individuals in a group synchronize their behavioural fluctuations (correlation) is possibly a more general mechanism than the synchronization of behaviour itself (order). All experimental studies performed up to now, however, concerned highly synchronized groups (as bird flocks, fish shoals and marching locusts \cite{cavagna+al_10, couzin_13, buhl+al_06}), which displayed both order and correlation. Hence, the question of whether or not order and correlations are two sides of the same phenomenon remained open until now. 
Here, we try and give an answer to this question by experimentally studying large swarms of insects in the field. As we will show, despite the lack of collective order, we do find strong correlations, indicating that even in biological systems collective behaviour and group-level coordination do not require order to be sustained.

\section*{Results}

{\bf Experiments and tracking.}
We perform an experimental study of swarms of wild midges in the field. Midges are small non-biting flies belonging to the order Diptera, suborder Nematocera (Diptera:Chiro\-no\-midae  and Diptera:Cera\-topo\-gonidae - see Methods). The body length of the species we study is in the range $1.2$--$2.4$mm. Swarms are found at sunset, in the urban parks of Rome, typically near stagnant water. As noted before \cite{downes_69}, we find that swarms form above natural or artificial landmarks. Moving the landmark leads to an overall displacement of the swarm. The swarms we studied range in size between $100$ and $600$ individuals (see Table S1).

To reconstruct the $3d$ trajectories of individual insects we use three synchronized cameras shooting at $170$ frames-per-seconds (trifocal technique -  Fig.\ref{fig:experiment}e and Methods). Our apparatus does not perturb the swarms in any way. 
The technique is similar to the one we used for starling flocks \cite{cavagna+al_08a}, with one notable difference. To reach the desired experimental accuracy we need to know the mutual geometric relations between the three cameras very accurately. In the case of flocks, this could be achieved only by an {\it a priori} alignment of the cameras. In the case of swarms, though, we proceed differently. After each swarm acquisition, we pin down the geometry of the camera system by taking multiple images of a calibrated target (Fig.\ref{fig:experiment}f). 
This procedure is so accurate that the error in the $3d$ reconstruction is dominated by the image segmentation error due to the pixel resolution. If we assume this to be equal to $1$ pixel (typically it is smaller than that because midges occupy many pixels), we make an error of $0.15$cm in the determination  of the distance between two points $5$cm apart from each other (a reference value for nearest neighbour distance). The absolute error is the same for more distant points, making the relative precision of our apparatus even higher. This accuracy makes the determination of the correlation functions we study here very reliable.
The $3d$-tracking of  each midge is performed by using the technique described in \cite{attanasi+al_13}.
Sample $3d$ reconstructions are shown in Fig.\ref{fig:experiment}b and in Video S1. Compared to previous field \cite{okubo_74,shinn+long_86,manoukis_09} and lab \cite{ouellette+al_13} studies, the present work is the most extensive experimental study of swarming insects in three dimensions to date. 

\begin{figure}[h!]
\includegraphics[width=1.0\columnwidth]{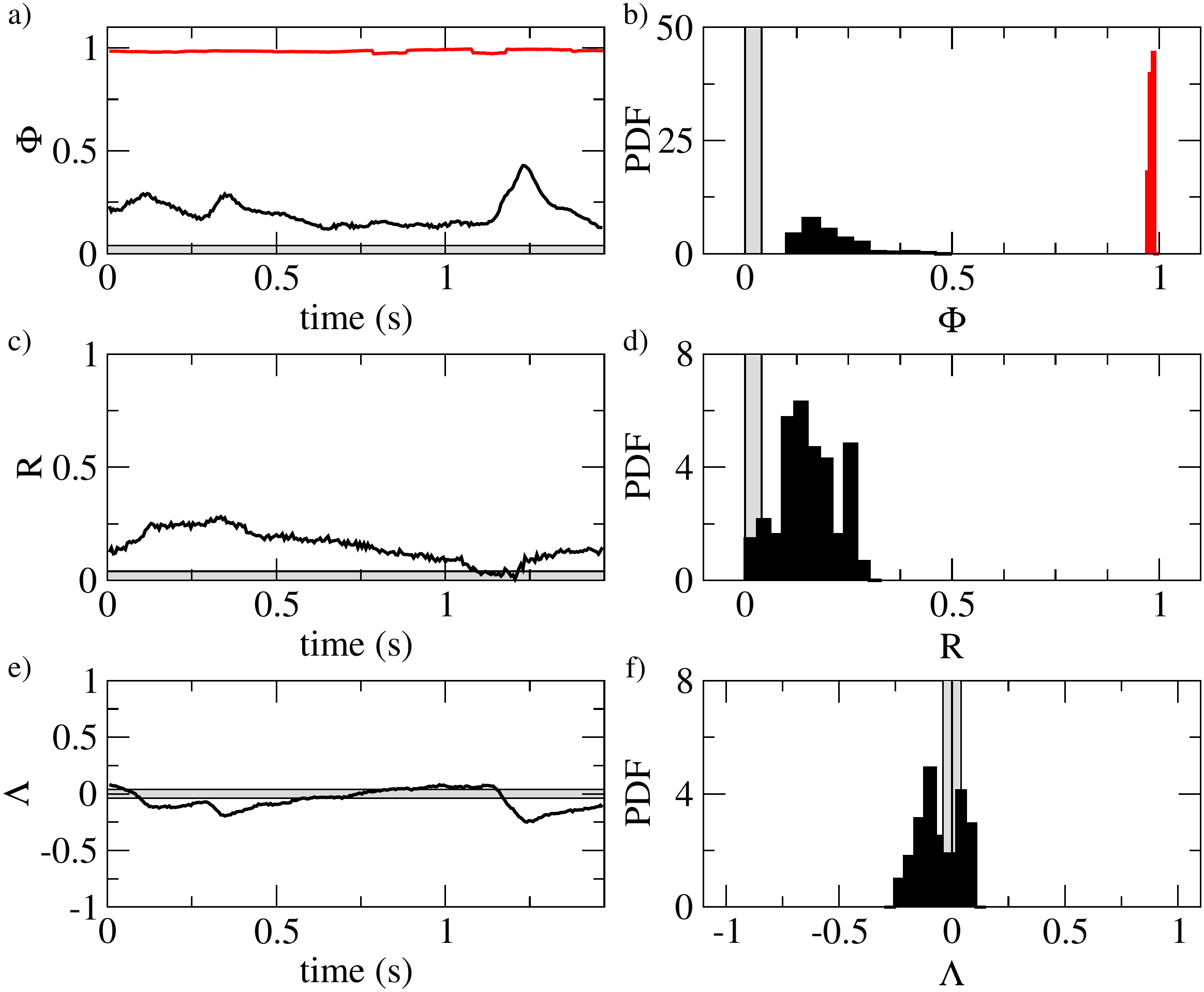}
\caption{
{\bf Natural swarms lack global order.}
Order parameters in a typical natural swarm.
In all panels the grey band around zero is
the expected amplitude of the fluctuations in a completely
uncorrelated system. In the left panels we report the time series of the order parameters, in the
right panels their probability distribution.
{\bf Top:} The alignment order parameter, known as polarization, $\Phi\in[0:1]$. In red we report the reference value of the polarization in a flock of starlings.
{\bf Middle:} Rotational order parameter, $R\in [0:1]$.
{\bf Bottom:} Dilatational order parameter $\Lambda\in[-1:1]$.
}
\label{fig:order}
\end{figure}

{\bf Lack of collective order.}
Swarms are in a disordered phase. The standard order parameter normally used in collective behaviour is the polarization, $\Phi = |(1/N) \sum_i \vec{v}_i/v_i |$, where $N$ is the number of midges in the swarm and $\vec{v}_i$ is the velocity of insect $i$. The polarization measures the degree of alignment of the directions of motion; it is a positive quantity and its maximum value is $1$. The average polarization over all swarms is quite small, $\Phi\sim0.21$ (see Fig.\ref{fig:order} and Table S1).  As a reference, in starling flocks we find $\Phi\sim 0.97$, on average \cite{cavagna+al_10}. 
The probability distributions of the polarization fully confirms the swarms' lack of translational order and the stark difference with flocks (Fig.\ref{fig:order}).
Clearly, swarms are not in a polarized state. Translation is not the only possible collective mode, though. For example, it is well-known that fish school can produce rotating (milling) configurations. Moreover, a group can expand/contract, giving rise to dilatational (or pulsive) collective modes. For this reason we have defined and measured also a rotational and a dilatational order parameter (see Methods). We find, however, that these quantities too have very small values (Fig.\ref{fig:order}). 
The time series, on the other hand, show that the order parameters can have rare, but strong fluctuations, during which their value may become significantly larger than that of an uncorrelated system (Fig.\ref{fig:order}). These large fluctuations are a first hint that nontrivial correlations are present.

\begin{figure}[h!]
\includegraphics[width=1.0\columnwidth]{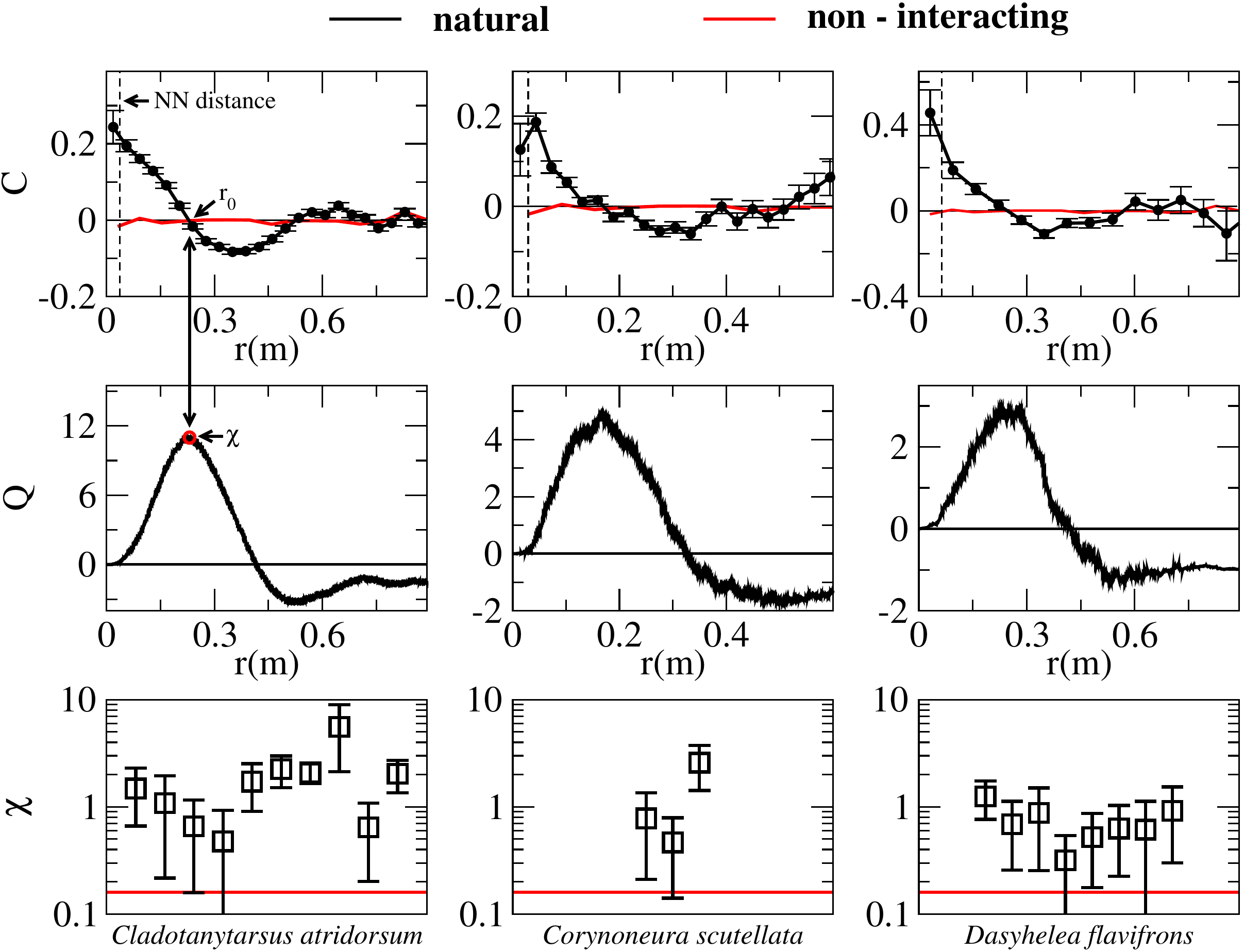}
\caption{
{\bf Swarms correlation.}
Black lines and symbols refer to natural swarms, red lines to simulations of `swarms'  of non-iteracting  particles (NHS). 
Each column refers to a different midge species.
{\bf Top}: Correlation function $C(r)$ as a function of the distance at one instant of time.
The dashed vertical line marks the average nearest neighbour distance, $r_1$, for that swarm. 
The correlation length, $r_0$, is the first zero of the correlation function. 
Red: correlation function in the NHS case. The value of $r$ for the NHS has been rescaled to 
appear on the same scale as natural distances. Each natural swarm is compared to a NHS 
with the same number of particles.
{\bf Middle:} Comulative correlation, $Q(r)$. 
This function reaches a maximum
$r=r_0$. The value of the integrated correlation at its maximum, $Q(r_0)$, is the
susceptibility $\chi$.
{\bf Bottom:} Numerical values of the susceptibility, $\chi$ in all analyzed swarms.
For each swarm the value of $\chi$ is a time average over the whole acquisition; error bars
are standard deviations. Red: the average susceptibility $\chi_\mathrm{NHS}$ in the noninteracting case.}
\label{fig:main}
\end{figure}

{\bf Correlation.}
The connected correlation function measures to what extent the change in behaviour of individual $i$  is correlated to that of individual $j$, at distance $r$. 
Correlation is the most accessible sign of the presence of interaction between the members of a group. The absence of interaction implies the absence of correlation. Conversely, the presence of correlation implies the presence of an effective interaction (see Supporting Information, Section I). 
Correlation can be measured for different quantities, but in the case of midges, as with birds and other moving animals, the principal quantity of interest is the direction of moti
To compute the connected correlation we first need to introduce the velocity fluctuations, namely the individual velocity subtracted of the overall motion of the group, $\delta \vec v_i = \vec v_i - \vec V_i$ (see Methods for the detailed definition of $\vec V_i$). This fluctuation is a dimensional quantity, hence it is unsuitable to compare the correlation in natural vs numerical systems, as we shall do later on. We therefore introduce the dimensionless velocity fluctuation,
\begin{equation}
\delta \vec \varphi_i = \frac{\delta \vec v_i}{\sqrt{ \frac{1}{N} \sum_k  \delta \vec v_k \cdot \delta\vec  v_k  }}  \ .
\label{pongo}
\end{equation}
The connected correlation function is then given by,
\begin{equation}
C(r)=\frac{\sum_{i\neq j}^N \ \vec{\delta \varphi_i} \cdot \vec{\delta \varphi_j}\  \delta(r-r_{ij})}{\sum_{i\neq j}^N \  \delta(r-r_{ij})} \ ,
\label{corr}
\end{equation}
where $\delta(r-r_{ij})=1$ if $r < r_{ij} < r+dr$ and zero otherwise, and $dr$ is the space binning factor. 
The form of $C(r)$ in natural swarms is reported in Fig.\ref{fig:main}: at short distances there is strong positive correlation, indicating that midges tend to align their velocity fluctuations to that of their neighbours. After some negative correlation at intermediate distances, $C(r)$ relaxes to no correlation for large distances.  This qualitative form is quite typical of all species analyzed (see  Fig.\ref{fig:main}). The smallest value of the distance where $C(r)$ crosses zero is the correlation length, $r_0$, that is an estimate of the length scale over which the velocity fluctuations are correlated \cite{cavagna+al_10}. The average value of this correlation length over all analyzed swarms is, $r_0  \sim 0.19$m. This value is about $4$ times larger than the nearest neighbours distance, whose average over all swarms is, $r_1 \sim 0.05$m (see Fig.\ref{fig:main} and Table S1). Previous works noticed the existence of pairing maneuvers and flight-path coordination between nearest neighbours insects \cite{okubo_74,okubo_86,butail+al_13}. Our results, however, indicate that midges within a natural swarm influence each other's motion far beyond their nearest neighbours.


{\bf Susceptibility.}
The collective response of the swarm depends crucially on two factors: how distant in space the behavioural change of one insect affects that of another insect
(spatial span of the correlation) and how strong this effect is (intensity of the correlation). To combine these two factors in one single observable 
we calculate the cumulative correlation up to scale $r$,
\begin{equation}
Q(r) = \frac{1}{N} \sum_{i\neq j}^N \ \vec{\delta \varphi_i} \cdot \vec{\delta \varphi_j}\  \theta(r-r_{ij})  \ .
\label{Q}
\end{equation}
It can be shown (see SI-II) that this dimensionless function is related to the space integral of the correlation function $C(r)$. 
Hence, $Q(r)$ reaches a maximum where $C(r)$ vanishes, i.e. for $r=r_0$ (see Fig.\ref{fig:main}). This maximum, $\chi\equiv Q(r_0)$, is a measure of the total amount of correlation present in the system. In statistical physics $\chi$ is exactly equal to the {\it susceptibility}, namely the response of the system to an external perturbation \cite{huang,binney+al_92}. In collective animal behaviour, we do not have a quantitative link between integrated correlation and response, so that calling $\chi$ susceptibility is not strictly correct.  However, if the probability distribution of the velocities is stationary, we can follow a maximum entropy approach \cite{bialek+al_12} and still find that the total amount of correlation in the system, $\chi$, is related to the way the group responds collectively to a perturbation (see SI-II). The value of $\chi$ for midges swarms is reported in Fig.\ref{fig:main}.

{\bf Noninteracting Swarm.}
In order to judge how significant is the correlation function $C(r)$ and how large is the susceptibility $\chi$ in natural swarms, we need an effective zero for these quantities, i.e. some null hypothesis baseline. As we have seen in the Introduction, the minimal assumption is that all individuals in the swarm interact with an external landmark independently from each other. Following Okubo \cite{okubo_86}
(but see also \cite{ouellette+al_13} and \cite{butail+al_13}), we therefore simulate a `swarm' of noninteracting particles performing a random walk in a three-dimensional harmonic potential (see Methods). Visually, the group behaviour of this Noninteracting Harmonic Swarm (NHS) looks remarkably similar to that of a real swarm (see Video S2 and S3): all `midges' fly around the marker and the group lacks collective order. 

This similarity, however, is deceptive. 
In the NHS, the correlation function $C(r)$ simply fluctuates around zero, with no spatial span, nor structure (Fig.\ref{fig:main}). 
Moreover, the susceptibility in the NHS is extremely small, $\chi_\mathrm{NHS} \sim 0.15$, whereas the susceptibility in natural swarms is up to $100$ times larger than this non-interacting benchmark (Fig.\ref{fig:main}). We conclude that swarming behaviour is {\it not} the mere epiphenomenon of the independent response of each insect with the marker.
Despite the lack of collective order, natural swarms are strongly correlated on large length scales. There are big clusters of midges that move coherently, contributing to the `dancing' visual effect of the swarm. The only way this can happen is that midges interact. What kind of interaction is that?

\begin{figure}[h!]
\includegraphics[width=1.0\columnwidth]{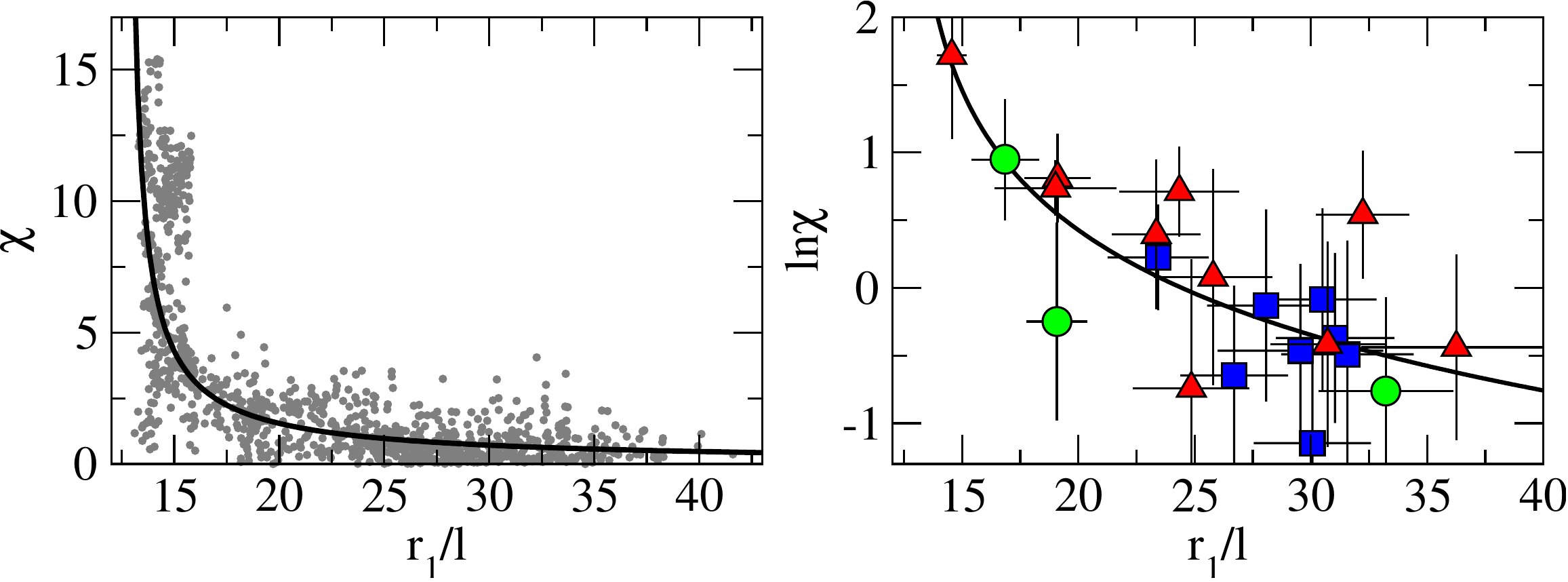}
\caption{
{\bf Swarms susceptibility.}
{\bf Left:} Susceptibility $\chi$ as a function of the rescaled nearest
neighbour distance, $r_1/l$, where $l$
is the body length. Each point represents a single time frame
of a swarming event, and all events are reported on the same
plot (symbols are the same for all species). 
The full line is the best fit to equation (\ref{umma}).
{\bf Right:} Logarithm of the average susceptibility as a function of
$r_1/l$. {\it Dasyhelea flavifrons} - blue squares; {\it Corynoneura scutellata} - green circles; {\it Cladotanytarsus  atridorsum} - red triangles.
 The full line represents the best fit to equation (\ref{umma}).
Each data point represents the time average over the entire acquisition of one
swarming event. Error bars indicate standard deviations. 
}
\label{fig:chi}
\end{figure}
%


{\bf Metric interaction.}
To answer this question we study the susceptibility across swarms of different densities. Interestingly, we find that $\chi$ increases when the average nearest neighbour distance, $r_1$, decreases (Fig.\ref{fig:chi}). Denser swarms are more correlated than sparser ones. This result indicates that midges interact through a metric perceptive apparatus: the strength of the perception decreases with the distance, so that when midges are further apart from each other (larger $r_1$) the interaction is weaker and the susceptibility $\chi$ is lower. This is at variance with what happens in bird flocks: birds interact with a fixed number of neighbours, irrespective of their nearest neighbour distance $r_1$ \cite{ballerini+al_08a}; such kind of topological interaction does not depend on the group density, hence the susceptibility does not depend on the nearest neighbour distance. Figure \ref{fig:chi}, on the other hand, shows that midges interact metrically, namely with all neighbours within a fixed metric range, $\lambda$. Hence, in swarms the number of interacting neighbours increases with decreasing $r_1$ (increasing density), and as a consequence of this increased amount interaction,  the system becomes also more correlated. 

In a system ruled by metric interaction we expect all lengths to be measured in units of the perception range, $\lambda$. This implies that the natural variable for the susceptibility is the rescaled nearest neighbour distance, $r_1/\lambda$. The problem is that we are considering different species of midges, likely to have different metric perception ranges. The simplest hypothesis we can make is that $\lambda$ is proportional to the insect body length $l$ (which we can measure), so that $\chi=\chi(r_1/l)$. This hypothesis is confirmed by the data: the susceptibility is significantly more correlated to the variable $r_1/l$ (P-value $=5\times 10^{-5}$) than to $r_1$ (P-value $=0.07$). The fact that the natural variable is $r_1/l$ is a further indication that the interaction in swarms is based on a metric mechanism.

The difference in the nature of the interaction  between flocking birds and swarming midges (topological vs. metric) is possibly due to the significant differences between vertebrates and arthropods. Topological interaction, namely tracking a fixed number of neighbours irrespective of their distance, requires a level of cognitive elaboration of the information \cite{ballerini+al_08a} more sophisticated than a metric interaction, where the decay of the effective force is merely ruled by the physical attenuation of the signal with increasing metric distance. 
In other words, within a metric mechanism the range of the interaction is fixed by a perceptive cut-off, rather than a cognitive one.
Metric interaction is known to be more fragile than topological one against external perturbations \cite{ballerini+al_08a}, and indeed it is far more likely to observe the dispersion of a swarm in the field than that of a flock. This may be the reason why the presence of an external marker is crucial for the swarming behaviour of midges \cite{downes_55}.

{\bf Correlation without order.}
The experimental observations of a notrivial connected correlation and of a large susceptibility indicate that midges are effectively interacting with each other by acting on their directions of motion. This does not exclude, of course, that other types of interaction are present.
First of all, the empirical observation that the swarm uses a visual marker as a reference for maintaining its mean spatial position, strongly suggests that each individual interacts with the marker. Besides, it is certainly possible that effective positional attraction-repulsion forces between midges, as those described in \cite{couzin+al_02}, exist.
However, the directional correlations indicate that insects are {\it also} effectively interacting by adjusting their velocities. Moreover, the fact that these correlations are positive for short distances means that midges tend to {\it align} their direction of motion. This fact may seem surprising, because alignment interactions typically lead to the formation of ordered (polarized)  groups, which is clearly not the case for midges. Swarms are disordered, and yet interacting and highly correlated systems. Is this a paradox?

In fact, it is not. An alignment interaction does not {\it per se} lead to global order in the group. In all models where imitation of the neighbours' is present, the onset of long-range order depends on the value of some key tuning parameter. In a ferromagnet, this parameter is the temperature $T$, namely the amount of noise affecting the interaction between the neighbouring spins. At high temperature the system is in a disordered state, whereas by lowering $T$ one reaches a critical temperature below which an ordering transition occurs. In models of active matter there is another parameter tuning the transition between disorder and order, that is density or, equivalently, nearest neighbour distance: the system gets ordered once the nearest neighbour distance falls below some transition value. The crucial point is that, in general, the correlation of the system tends to be very large {\it around} the transition point, irrespective of whether the system is in the ordered or in the disordered phase. Hence, even a disordered system can display large correlations, provided that it is not too far from an ordering transition. In what follows, we want to show that this is indeed what happens with midge swarms.

\begin{figure}[h!]
\includegraphics[width=1.0\columnwidth]{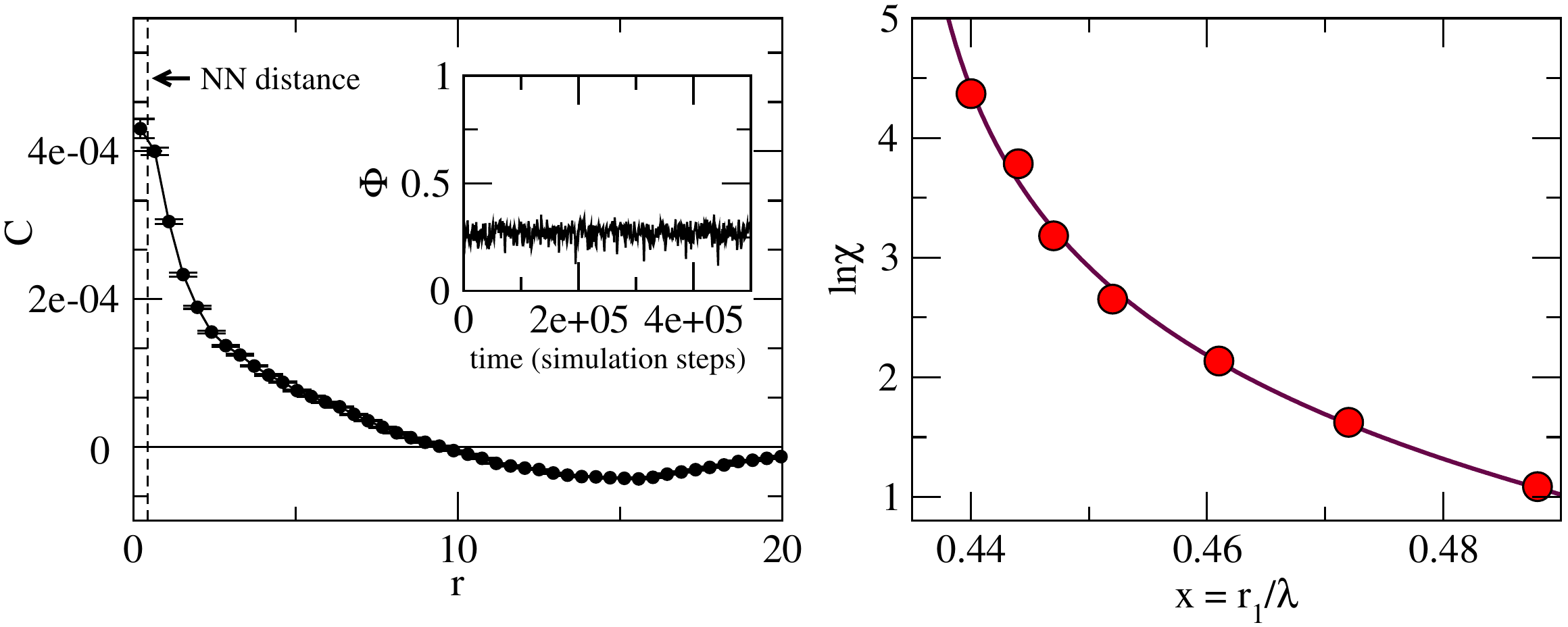}
\caption{
{\bf Vicsek model.} 
Three-dimensional Vicsek model in a central potential.
{\bf Left:} Correlation function $C(r)$ in the disordered phase, but close to the ordering transition. The dashed line is the nearest neighbour distance.
Inset: polarization as a function of time. For this value of $x$ the system
is disordered.
{\bf Right:}  Logarithm of the susceptibility as a function of the rescaled nearest neighbour distance, $x=r_1/\lambda$, where $\lambda$ is the metric interaction range.
The full line represents the 
best fit to equation (\ref{umma}).}
\label{fig:vicsek}
\end{figure}

{\bf Vicsek model.}
The simplest model based on alignment interaction that predicts an order-disorder transition on changing the density is the Vicsek model of collective motion \cite{vicsek+al_95}.
In this model each individual tends to align its direction of motion to that of the neighbours within a {\it metric} perception range, $\lambda$.  The rescaled nearest neighbour distance, $x\equiv r_1/\lambda$, is the control parameter: for low noise, the model predicts a transition from a disordered phase (low polarization) at high values of $x$ (low density), to an ordered phase (large polarization) at low values of $x$ (high density) \cite{vicsek+al_95, gonci+al_08, chate+al_08}. 
We numerically study the Vicsek model in three dimensions. As we have seen, real swarms hold their average position with respect to a marker; to reproduce this behavioural trait we introduce an harmonic attraction force that each individual experiences towards the origin (see Methods). Also in central potential the model displays an ordering transition: at large density, for $x<x_c$ the system is ordered and it has large polarization (Video XXXX). On the other hand, the polarization is low in the disordered phase, $x>x_c$ (Fig.\ref{fig:vicsek}). However, the correlation function is nontrivial when $x$ is sufficiently close to $x_c$ (Fig.\ref{fig:vicsek}), indicating the existence of large clusters of correlated individuals, which can be clearly detected in Video XXXX. We calculate the susceptibility $\chi$, in the same manner as we did for natural swarms, in the disordered phase, $x>x_c$ and find a clear increase of $\chi$ on lowering $x$ (Fig.\ref{fig:vicsek}). 

This increase of the susceptibility is coherent with the existence of an ordering transition at $x_c$.
In the thermodynamic limit, the transition in the metric Vicsek model is first order \cite{chate+al_08}. However,  it has been shown that, unless $N$ is much larger than the values analyzed here, the transition is hardly distinguishable from a second order one \cite{vicsek+al_95,chate+al_08}. As a consequence, 
the susceptibility is expected to become very large approaching $x_c$ and to follow the usual scaling relation of critical phenomena \cite{baglietto+albano_08},
\begin{equation}
\chi \sim \frac{1}{(x - x_c)^\gamma}  \quad , \quad x=r_1/\lambda \ .
\label{umma}
\end{equation}
A fit to equation (\ref{umma}) of the $3d$-Vicsek data is reported in Fig.\ref{fig:vicsek}, giving $\gamma = 1.5 \pm 0.1 $ and a transition point, $x_c=0.434$. The reason for the growth of $\chi$ approaching $x_c$ in the Vicsek model is quite intuitive. The model is metric, so that at large $x$, namely when the nearest neighbour distance $r_1$ is much larger than the interaction range $\lambda$, very few individuals interact with each other, and coordination is small. The smaller $x$ becomes, the larger the number of particles within the mutual interaction range, thus promoting the correlation of larger and larger clusters of particles. For this reason the correlation length and the susceptibility grow when the nearest neighbour distance decreases. When $x$ approaches its critical value, the coordinated clusters become as large as the whole system, so that the groups orders below $x_c$.

%

The low order parameter, the nontrivial correlation function, and especially the increase of $\chi$ on decreasing the nearest neighbour distance, are phenomenological traits that the metric Visek model shares with natural swarms. We conclude that a system based solely on alignment can be in its disordered phase and yet display large correlations, as midge swarms do. 
It is interesting to note that by approaching the ordering transition a compound amplification of the coorelation occurs: when the nearest neighbour distance, $r_1$, decreases, the spatial span of the correlation, $r_0$, increases, so that the effective perception range in units of nearest neighbour distance, $r_0/r_1$ is boosted.
We emphasize that we are not quantitatively fitting Vicsek model to our data. Our only aim is to demonstrate a general concept: large correlation and lack of global order can coexist even in the simplest model of nearest neighbours alignment, provided that the system is sufficiently close to an ordering transition.

{\bf Estimating the interaction range.}
The consistency between our experimental data and the Vicsek model suggests that an underlying ordering transition could be present in swarms as well.  An ordering transition as a function of the density has been indeed observed in laboratory experiments on locusts \cite{buhl+al_06}, fish \cite{becco_02} and in observations of oceanic fish shoals \cite{makris+al_09}. In these cases, both sides (low and high density) of the ordering transition were explored. However, midge swarms in the field are always disordered, living in the low-density/high-$x$ side of the transition. Locating a transition point having data on just one side of it, is a risky business. The reason why we want to do this here is because it will allow us to give a rough estimate of the metric range of interaction in midges, which can be compared with other experiments. 

If a Vicsek-like ordering transition exist, we can use equation (\ref{umma}) to fit the swarms data for $\chi$ (Fig.\ref{fig:chi}). As we already mentioned, 
we do not know the value of the metric perception range, $\lambda$, in swarms. Therefore, we use as scaling variable $r_1/l$, where $l$ is the body length. 
Although the fit works reasonably well (Fig.\ref{fig:chi}), the scatter in the data is quite large; hence, given the nonlinear nature of the fit, it would be unwise 
to pin down just one value for the parameters, and we rather report confidence intervals. The fit gives a transition point in the range, $(r_1/l)_c \in [9.0:13.5]$, 
with an exponent in the range, $\gamma\in[0.75:1.3]$ (larger exponents correspond to lower transition points).

\begin{figure}[h!]
\includegraphics[width=0.8 \columnwidth]{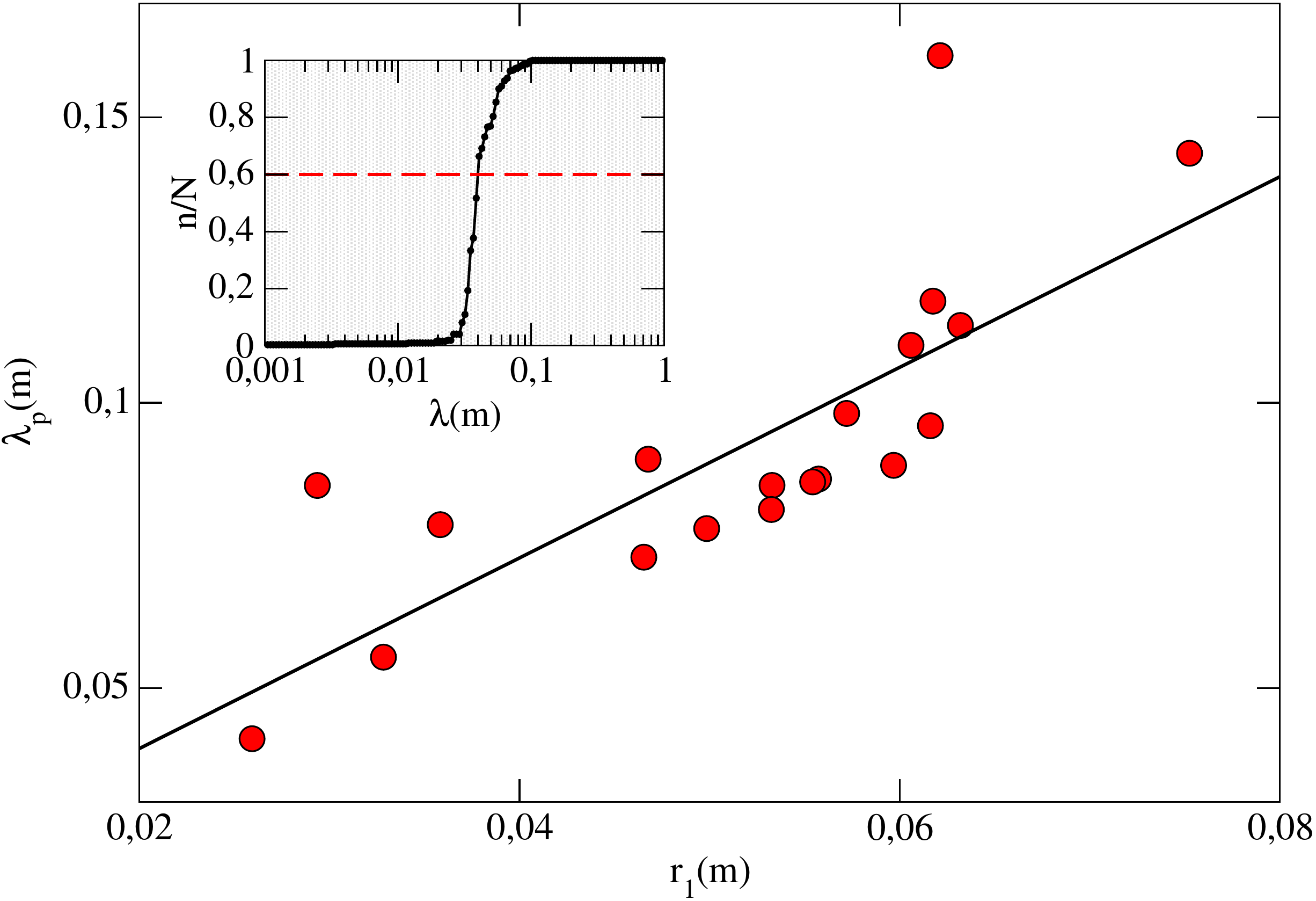}
\caption{{\bf Percolation threshold.}
Percolation threshold $\lambda_p$ as a function of the nearest-neighbor distance in natural swarms. The linear fit (black line) gives, $\lambda_p = 1.67 r_1$.
Inset:  Fraction of midges belonging to the largest cluster as a function of the clustering threshold $\lambda$. In correspondence of the percolation threshold $\lambda_b$ there is the formation of a giant cluster. We define $\lambda_p$ as the point where $n/N=0.6$ (red dashed line). Because of the sharp nature of the percolation transition, the value of $\lambda_p$ does not depend greatly on the threshold used.}
\label{fig:percolation}
\end{figure}

Interestingly, there is an alternative way to locate the ordering transition that does not rely on the fit of $\chi$. Let us establish a link between pairs of insects closer than the perception range $\lambda$ and calculate the size of the biggest connected cluster in the network. Given a swarm with nearest neighbour distance $r_1$, the larger $\lambda$, the larger this cluster. When $\lambda$ exceeds the percolation threshold, $\lambda>\lambda_p(r_1)$, a giant cluster of the same order as the group size appears \cite{percolation}. 
We calculate the percolation threshold in swarms (Fig.\ref{fig:percolation} and Methods) and find $\lambda_p = 1.67\, r_1$. 
The crucial point is that varying the perception range $\lambda$ at fixed nearest neighbour distance $r_1$, is equivalent to varying $r_1$ at fixed $\lambda$.
Hence, at fixed $\lambda$, there is an equivalent percolation threshold of the nearest neighbour distance, $(r_1)_p$, such that for $r_1 < (r_1)_p$ a giant cluster appears. Clearly, 
$(r_1)_p \sim \lambda/1.67=0.6 \lambda$. 
It is reasonable to hypothesize that the critical nearest neighbour distance is close to the maximal distance compatible with a connected network, given $\lambda$. A sparser network would cause the swarm to lose bulk connectivity. 
Therefore, given a certain perception range $\lambda$, the ordering transition occurs at values of the nearest neighbour distance $r_1$ close to its percolation threshold, $(r_1/\lambda)_c \sim (r_1)_p /\lambda \sim 0.6$. 

At this point we have two independent (and possibly equally unreliable) estimates of the transition point in natural swarms of midges: 
the first one in units of body-lenghts, $(r_1/l)_c \in [9.0:13.5]$; the second one in units of interaction range, $(r_1/\lambda)_c \sim 0.6$. Putting the two together we finally obtain an estimate of the metric interaction range in units of body-lengths, $\lambda \sim 15-22\,l$.
The body length of the species under consideration is in the range,  $l\sim 1.2$mm$-2.4$mm. This implies a perception range of a few centimeters, $\lambda \sim 2-5$cm, depending on the species.
This crude estimate of the midge interaction range is compatible with the hypothesis that midges interact acoustically. In \cite{fyodorova+al_03} the male-to-male auditory response in {\it Chironomus annularius} (Diptera:Chironomidae) was studied and it was found that the range of the response was about $1.0-1.5$cm, not too far from our estimate. Similar measurements in mosquitoes (Diptera:Culicidae) show that the auditory perception range is about $2$cm \cite{gibson_10}, which is again compatible with our determination of the interaction range in midge swarms.

\section*{Discussion}

We have shown that natural swarms of midges lack collective order and yet display strong correlations. Such correlations extends spatially much beyond the inter-individual distance, indicating the presence of significant cluster of coordinated individuals.
This phenomenology is  incompatible with a system of noninteracting particles whose swarming behaviour is solely due to the attraction to an external landmark. 
We conclude that genuine collective behaviour is present in swarms.
By using Vicsek model as a simple conceptual framework, we have showed that this coexistence of disorder and correlation is a general feature of systems with alignment interaction close to their ordering transition. 

We stress that the existence of correlation, and therefore of inter-individual interaction, is not in contradiction with the fact that a swarm almost invariably forms in proximity of a marker. The effect of the marker (external force) is merely to keep the swarm at a stationary position with respect to the environment. However, as we have shown in the case of the noninteracting swarm, this stationarity (which superficially would seems the only visible trait of swarming), cannot by itself produce the observed strong correlations. 

We believe that correlation, rather than order, is the most significant experimental  signature of collective behaviour.
Correlation is a measure of how much and how far the behavioural change of one individual affects that of other individuals not directly interacting with it.
Our data show that in swarms correlations are so strong that the effective perception range of each midge is much larger than the actual interaction range. 
If the change of behaviour is due to some environmental perturbations, such large 
correlation guarantees that the stimulus is perceived at a collective level. 

A notion of collective behaviour based on correlation is more general and unifying than one based on order. 
For example, bird flocks and insect swarms look like completely different systems as long as we stick to collective order. However, once we switch to correlation, we understand that  this big difference may be deceptive: both flocks and swarms are strongly correlated systems, in which the effective perception range, or correlation length, is far larger than the interaction range \cite{cavagna+al_10}. In this perspective, the striking difference in emergent order between the two systems, namely the fact that flocks move around the sky, whereas swarms do not, may be related to different ecological factors, rather than to any fundamental qualitative difference in the way these systems interact. Strong correlations similar to those found in bird flocks and midge swarms  have also been experimentally measured in neural assemblies \cite{schneidman_bialek_2006}. This huge diversity - birds, insects, neurons - is bewildering but fascinating, and it suggests that correlation may be a universal condition for collective behaviour, bridging the gap between vastly different biological systems.


\vskip 0.3 truecm
{\bf Acknowledgments.}
We thank William Bialek, Enzo Branchini, Massimo Cencini, Francesco Ginelli and Dov Levine for discussions. We also acknowledge the help of Tomas S. Grigera in running numerical tests on the single sample susceptibility in the Ising model. This work was supported by grants IIT--Seed Artswarm, ERC--StG n.257126 and US-AFOSR - FA95501010250 (through the University of Maryland).
\vskip 0.3 truecm

{\bf Authors Contribution.}
A.C. and I.G designed the study. A.C. coordinated the experiment. A.A., A.C., L.D.C., I.G., S.M., L.P., E.Shen and M.V. set up and calibrated the 3d system. L.D.C., S.M., O.P. and E.Shen performed the experiment. B.R. performed the species recognition and monitored all entomological aspects of the study. A.A., L.D.C., S.M., L.P., E.Shen, E.S. and M.V. performed the tracking and produced the $3d$ data. A.C., L.D.C., I.G., S.M., L.P. and M.V. analyzed the data.  E.S. performed the numerical simulations. A.C. wrote the paper. Correspondence and requests for materials should be addressed to S.M. (stefania.melillo79@gmail.com) or A.C. (andrea.cavagna@roma1.infn.it).



\section*{Methods}
\label{methods}

{\bf Experiments.}
Data were collected in the field (urban parks of Rome), between May and October, in $2011$ and in $2012$.
We acquired video sequences using a multi-camera system of three synchronized cameras (IDT-M5) shooting at $170$ fps. Two cameras (the stereometric pair) were at a distance between $3\mathrm{m}$ and $6\mathrm{m}$ depending on the swarm and on the environmental constraints. A third camera, placed at a distance of $25\mathrm{cm}$ from the first camera was used to solve tracking ambiguities. We used Schneider Xenoplan $50\mathrm{mm}$~$f/2.0$ lenses. Typical exposure parameters: aperture $f/5.6$, exposure time $3$ms. Recorded events have a time duration between $1.5$ and $15.8$ seconds. No artificial light was used. To reconstruct the $3d$ positions and velocities of individual midges we used the techniques developed in \cite{attanasi+al_13}. Wind speed was recorded. After each acquisition we captured several midges in the recorded swarm for lab analysis. A summary of all swarms data can be found in Table S1.

{\bf Midge identification.}
We recorded swarms of midges belonging to the family Diptera:Ceratopogonidae ({\it Dasyhelea flavifrons}) and Diptera:Chironomidae ({\it Corynoneura scutellata} and {\it Cladotanytarsus  atridorsum}). 
Midges belonging to the family Chironomidae were identified to species  according to \cite{Langton_07}, 
the ones belonging to the family Ceratopogonidae were identified according to \cite{Kieffer_25} and \cite{Dominiak_12}.
Specimens used for identification were captured with a hand net and fixed in  70$^\circ$ alcohol,  cleared and prepared according to 
\cite{Wirth_68}. Permanent slides were mounted in Canada Balsam and dissected according to \cite{Wiederholm_89}.   
Species identification was based on morphology of the adult male, considering different characters, as wing venation, antennal ratio
(length of apical flagellomere divided by the combined length of the more basal flagellomeres) and
genitalia, which in Diptera are named hypopygium (a modified ninth abdominal segment together with the copulatory apparatus - see Fig.\ref{fig:experiment}).

{\bf Definition of the velocity fluctuations.}
Let $\{\vec{x}_i(t)\}$ and $\{\vec{x}_i(t+\Delta t)\}$ be respectively the sets of coordinates of the objects in the system at 
time $t$ and $t+\Delta t$. For the sake of simplicity in the notation, assume $\Delta t=1$.
We define the velocity vector of each object $i$ as $\vec{v}_i(t)=\vec{x}_i(t+1)-\vec{x}_i(t)$.

To compute the correlation function we need to subtract the contributions due to the collective motion from the velocity. 
We identify three collective modes: translation, rotation and dilatation (expansion/contraction).

Let $\vec{x}_0(t)=1/N\sum_i\vec{x}_i(t)$ be the position of the center of mass at time $t$ and 
let $\vec{y}_i(t)=\vec{x}_i(t)-\vec{x}_0(t)$ denote the position of the $i$-th object in the center 
of mass reference frame. We define the optimal 
translation as the displacement $\Delta\vec{x}_0(t)=\vec{x}_0(t+1)-\vec{x}_0(t)$.

The velocity fluctuation subtracting the overall translation is then $\delta\vec{v}_i=\vec{y}_i(t+1)-\vec{y}_i(t)$.

The optimal rotation about the origin is defined as the $3x3$ orthogonal matrix $\bf{R}$ which minimizes the quantity 
$\sum_i[\vec{y}_i(t+1)-{\bf R}\vec{y}_i(t)]^2$, see \cite{kabsch_76}.
The velocity fluctuations subtracting the optimal translation and rotation is $\delta\vec{v}_i=\vec{y}_i(t+1)-{\bf R}\vec{y}_i(t)$.

Finally the optimal dilatation is defined as the scalar ${\bf\Lambda}$ that minimizes the quantity
$\sum_i[\vec{y}(t+1)-{\bf \Lambda R}\vec{y}_i(t)]^2$, see \cite{kabsch_76}.

Therefore the velocity fluctuation subtracting the optimal translation, rotation and dilatation is:
\begin{equation}\label{zebra}
\delta\vec v_i =\vec{y}_i(t+1)-{\bf\Lambda R}\vec{y}_i(t)\equiv \vec v_i -\vec V_i \ .
\end{equation}
where $\vec{V}_i$ is the contribution to the individual velocity due to the three collective modes of the system.

{\bf Rotation and dilatation order parameters.}
The rotational order parameter is defined as,
\begin{equation}
R=\frac{1}{N}\left|\sum_i \frac{\vec y_i^\bot(t)\times \vec v_i(t)}{|\vec y_i^\bot(t)\times \vec v_i(t)|}\cdot\hat K\right|  \ ,
\label{rot_param}
\end{equation}
where $\vec y_i^\bot$ is the projection of $\vec y_i(t)$ on the plane orthogonal to the axis of rotation, the cross indicates a vectorial product, and $\hat K$ is a unit vector in the direction of the axis of rotation. In (\ref{rot_param}), $\vec y_i^\bot(t)\times \vec v_i(t)$ is the angular momentum of midge $i$ with respect to the
axis $\hat K$. In a perfectly coherent rotation all individuals would have angular momenta parallel to the axis, so that $R=1$. In a noncoherent system, some
of the projections of the angular momentum on $\hat K$ would be positive and some negative, so $R\sim 0$. 
Note that $\hat K$ is the axis of the rotation defined in the previous section, computed using kabsch algorithm \cite{kabsch_76}.

The dilatational order parameter is defined as,
\begin{equation}\label{eqn::lambda}
\Lambda = \frac{1}{N}\sum_i 
\frac{[R \, \vec y_i(t)] \cdot [\vec y_i(t+\Delta t)-R \, \vec y_i(t) ]}{|R \,\vec y_i(t) |\; |\vec y_i(t+\Delta t)-R\,\vec y_i(t)|}
\ . 
\end{equation}
$\Lambda\in[-1,1]$ and it measures the degree of coherent expansion (positive $\Lambda$) and contraction (negative $\Lambda$) of the swarm.
In a perfectly coherent expansion/contraction $\vec{y}_i(t+\Delta t)-{\bf R}\vec{y}_i(t)$ would be parallel to ${\bf R}\vec{y}_i(t)$ and so the scalar product in equation \ref{eqn::lambda} 
will be $1$ for an expansion and $-1$ for a contraction.

{\bf Normalization of the correlation function}. 
In the study of flocks \cite{cavagna+al_10}, we normalized $C(r)$ by its limiting value for $r\to0$, which is equivalent to divide it by the value in the first bin. In that way the normalized correlation function tends to $1$ for $r\to0$, so that its value is amplified. In the study of flocks we were only looking at the correlation length, which is not altered by such a normalization. However, here we will be interested in both the range and the intensity of the correlation, so we must not amplify artificially the correlation signal. 
Normalizing the fluctuations as in (\ref{pongo}) is equivalent to normalize the correlation function by its value at {\it exactly} $r=0$, i.e. for $i=j$, which is different from its limit for $r\to 0$.

{\bf Noninteracting Harmonic Swarm.}
The NHS is an elementary model of non interacting particles performing a random walk in a three-dimensional harmonic potential. The dynamics of each particle is defined by the Langevin equation,
\begin{equation}
\label{langevinNHS}
m\ddot{\vec{x}}_i(t)=-\gamma\dot{\vec{x}}_i(t)-k\vec{x}_i(t)+\sqrt{\eta \gamma}\vec{\xi}_i(t) \ ,
\end{equation}
where $\vec{x}_i(t)$ is the position of  the $i$-th particle at time $t$, $m$ is the mass, $\gamma$ the friction coefficient, $k$ the harmonic constant and $\vec{\xi}_i(t)$ is a random vector with zero mean and unit variance,  $\langle \xi_i^\alpha(t)\xi_j^\beta(t^\prime)\rangle =\delta(t-t^\prime) \delta_{i,j}\delta_{\alpha,\beta}$, with $\alpha=x,y,z$. Clearly, in this model there is no interaction between particles. The parameter $\eta$ tunes the strength of the noise. The equation of motion are integrated with the Euler method \cite{butcher_03}. We simulated the NHS in the critically damped regime ($\gamma^2 = 4 m k$), which gives the best similarity to natural swarms. The number of particles $N$ is set equal to that of the natural swarm we want to compare with. Parameters have been tuned to have a ratio between the distance traveled by a particle in one time step (frame) and the nearest neighbor distance comparable to natural swarms, $\Delta r/r_1 \sim 0.15$: $m=1, k=12.75, \gamma=7.14, \eta=2.0$.

{\bf Vicsek model.}
We performed numerical simulations of the Vicsek model in $3d$ \cite{vicsek+al_95,czirok+al_99,gonci+al_08,chate+al_08,baglietto+albano_08}. The direction of particle $i$ at time $t+1$ is the average direction of all particles within a sphere of radius $\lambda$ around $i$ (including $i$ itself). The parameter $\lambda$ is the metric radius of interaction. The resulting direction of motion is then perturbed with a random rotation (noise).  
Natural swarms are known to form close to a marker and to keep a stationary position with respect to it \cite{downes_55}. 
To mimic this behaviour we modified the Vicsek model by adding an external harmonic force equal for all particles. This potential also grants cohesion, without the need to introduce an inter-individual attraction force \cite{okubo_86,ouellette+al_13,butail+al_13}.
The update equation for velocities is therefore given by,
\begin{equation}
\label{viccent}
 \vec{v}_i(t+1)=v_0 \ \mathcal{R}_\eta \left[  \Theta \left( \sum_{j\in S_i} \vec{v}_j(t) - \beta\vec{r}_i(t)\right) \right] \ ,
\end{equation}
where $S_i$ is the spherical neighborhood of radius $\lambda$ centered around $i$,  $\Theta$ is the normalization operator, $\Theta(\vec{x})=\vec{x}/|\vec{x}|$, and $\mathcal{R}_\eta$ performs a random rotation uniformly distributed around the argument vector with maximum amplitude of $4\pi\eta$.
The term $-\beta\vec{r}_i(t)$ is the harmonic force directed towards the origin. For $\beta=0$ we recover the standard Vicsek model. The update equation for the positions is, $\vec{r}_i(t+1)=\vec{r}_i(t)+\vec{v}_i(t+1)$. Thanks to the central force we can use open boundary conditions. Particles have all fixed velocity modulus $|\vec{v}|=v_0=0.05$. Each simulation has a duration  of $6\times10^5$ time steps, with initial conditions consisting in uniformly distributed positions in a sphere and uniformly distributed directions in the $4 \pi$ solid angle. After a transient of $10^5$ time steps, we saved 500 configurations at intervals of 1000 time steps in order to have configurations with velocity fluctuations uncorrelated in time. 
The control parameter of interest is $x\equiv r_1/\lambda$, where $r_1$ is the nearest neighbour distance, which is tuned by $\beta$. The model displays a transition to an ordered phase when $x<x_c$. We studied the susceptibility $\chi(x)$ for different values of $x\in [0.34,0.70]$. To observe the power-law behaviour of $\chi(x)$ 
predicted by the model we performed standard finite-size scaling  \cite{baglietto+albano_08}: at each fixed value of the system's size $N\in[128,8192]$ we calculated $\chi(x;N)$ and worked out the maximum of the susceptibility $\chi_\mathrm{max}(N)$ and its position $x_\mathrm{max}(N)$; we finally plotted $\chi_\mathrm{max}$ vs. $x_\mathrm{max}$ parametrically in $N$, to obtain the function $\chi(x)$ in Fig.\ref{fig:vicsek}. The noise, $\eta$, affects the position of the transition point point $x_c$ \cite{vicsek+al_95,gonci+al_08,chate+al_08}, but this is irrelevant for us, because we do not use  any quantitative result from the model to infer the biological parameters of real swarms.  
The data reported in Figs.\ref{fig:vicsek} have $\eta=0.45$.

{\bf Percolation threshold.}
For each frame we run a clustering algorithm with scale $\lambda$ \cite{lu_78}: two points are connected when their distance is lower than $\lambda$.
For each value of $\lambda$ we compute the ratio $n/N$ between the number of objects in the largest cluster and the total number of objects in the swarm (Fig.\ref{fig:percolation}). The percolation threshold, $\lambda_c$, is defined as the point where a giant cluster, i.e. a cluster with size of the same order as the entire system, forms \cite{percolation}. We define $\lambda_c$ as the point where $n/N=0.6$. The percolation threshold scales with the nearest neighbour distance, $\lambda_c = 1.67 \,r_1$ (Fig.\ref{fig:percolation}).
Strictly speaking, the percolation argument only holds at equilibrium, because in a system where particles are self-propelled there may be order even at low density \cite{gonci+al_08}. However, at low values of the noise, we still expect the percolation argument to give a reasonable, albeit crude, estimate of the perception range.

\section*{Supplementary Information}

\section{Connected vs nonconnected correlation}

The most basic kind of correlation one can measure in a system is the scalar product of the velocities of two different individuals, $\vec v_i\cdot\vec v_j$. This quantity is large if velocities are pointing in the same direction and low if they are uncorrelated. This is what is called {\it non-connected} correlation, and it has a problem: its value is trivially dominated by the {\it mean} motion of the system. Imagine that a gust of wind shifts the entire swarm, so that each midges's velocity is dominated by the wind speed. As an effect of the wind, midges $i$ and $j$ would be moving nearly parallel to each other, so that the non-connected correlation would be high. This, however, is simply an effect of the wind, and it is not a genuine sign of correlation, nor of interaction between the individuals. The same thing would happen in a system of uncorrelated particles put in rotational motion around an axis: velocities of nearby particle are mostly parallel as a mere effect of the overall rotation. 

These examples show that, in order to get information about the {\it bona fide} interaction between individuals,  we need to compute the correlation between the {\it fluctuations} around the mean motion of the system. In other words, what we want to detect is to what extent the individual changes of behaviour with respect to the global behaviour of the system are correlated. This is what the {\it connected} correlation does and it is the only reliable measure of correlation in a system. The presence of a non-connected correlation is not in general proof of anything at the level of the interaction, as the wind example clearly shows. On the other hand, the presence of  non-zero connected correlation in a system is unambiguous proof that there is interaction, and strong enough to produce collective effects. 

To compute the connected correlation function we therefore need to identify the collective modes of motion of the system and subtract them from the individual motion (see Methods). In this way we obtain the velocity fluctuation, namely the velocity of midge $i$ in a reference frame that not only is co-moving with the centre of mass, but also rotating and expanding/contracting as the whole swarm. Therefore, what is left is the deviation of $i$ from the mean group motion, which is the only quantity that is safe to correlate. 

It is very important to realize that an error or an artifact in computing the fluctuations can lead to spurious values of the correlation. As an example, consider two different and unrelated swarms moving in opposite directions, because of some weird fluctuation of the wind. If we fail to notice that these are {\it two} systems and analyze our data as if they were {\it one}, we get a zero net motion of the centre of mass. Hence, the velocity fluctuations are equal to the full velocities, and we are effectively computing a non-connected correlation, rather than a connected one, giving the delusion of very large correlation.

In the main text we show that swarms are mostly disordered. However, the fact that order parameters are low on average, does not mean that we can use the full velocities to compute the correlation function. As we have already said, a brief gust of wind can push the non-connected correlation function, to very high values. In this study, we are not investigating the origin of the order parameters fluctuations, but we focus on correlations. Hence, we have to be sure that correlation is computed in a way to avoid any spurious signal from the collective modes.

%
%

\section{Susceptibility, response and correlation}
\label{spin}
In a stationary system, it can be proved \cite{binney+al_92} that the susceptibility is equal to the collective response of the system to uniform external perturbations. Maximum entropy calculations \cite{bialek+al_12} show that the stationary probability distribution of the velocities in systems where there is an alignment interaction is given by,
\begin{equation}
P(v) = \frac{1}{Z} \; e^{J \sum_{i,j} \vec v_i \cdot \vec v_j}  \ ,
\end{equation}
where $J$ is the strength of the interaction (depending on distance $r$ in a metric system) and $Z$ is a normalizing factor (the partition function),
\begin{equation}
Z=\int Dv  \; e^{J \sum_{i,j} \vec v_i \cdot \vec v_j} \ .
\end{equation}
If an external perturbation (or field) $h$ couples uniformly to all velocities, this distribution gets modified as, 
\begin{equation}
P(v) = \frac{1}{Z(h)}\; e^{J \sum_{i,j} \vec v_i \cdot \vec v_j + \vec h\cdot \sum_i \vec v_i} \ .
\end{equation}
Now we ask what is the collective response $\chi$ of the system to a small variation of the perturbation $h$. To answer this question we calculate the variation of the global order parameter, i.e. of the space average of the velocity, under a small variation of $h$. we have,
\begin{eqnarray}
\chi &=& \frac{\partial}{\partial h}  \langle \frac{1}{N}\sum_k  v_k \rangle 
\nonumber
\\
&=&\frac{\partial}{\partial h}  \int Dv \;  P(v) \frac{1}{N}\sum_k  v_k 
\nonumber
\\
&=& \frac{1}{N}\sum_{i,k} \int Dv P(v)  v_k  v_i  
- \int Dv P(v) v_i \int Dv P(v)  v_k 
\nonumber
\\
&=&  \frac{1}{N}\sum_{i,k} \langle v_k  v_i \rangle -  \langle v_i \rangle \langle   v_k \rangle = \frac{1}{N}\sum_{i,k} \langle \delta v_k \; \delta v_i \rangle \ ,
\label{monster}
\end{eqnarray}
where we have disregarded the vectorial nature of the quantities not to burden the notation and where we have defined,
\begin{equation}
\langle f(v) \rangle = \int Dv P(v) f(v) \ .
\end{equation}
Apart from the missing normalization, needed to make $\chi$ dimensionless, the quantity in (\ref{monster}) is exactly the susceptibility defined in the main text, equation $(2)$.

Let us now analyze in detail the relation between correlation function and susceptibility in a finite size system, where instead of the ensemble averages, $\langle\cdot\rangle$, we can only perform space averages. From equations $(1)$ and $(2)$ in main text, we obtain:
\begin{equation}
Q(r) = \frac{1}{N}\int_0^r dr' \sum_{i\neq j}^N \  \delta(r'-r_{ij}) \, C(r')  \ .
\end{equation} 
If we make the hypothesis that mass fluctuations are not strong, we can write,
\begin{equation}
\frac{1}{N} \sum_{i\neq j}^N \  \delta(r'-r_{ij}) \sim 4\pi x^2 \rho  \ ,
\end{equation}
where $\rho$ is the density. Hence, we get,
\begin{equation}
Q(r) =  \frac{3}{r_1^3}   \int_0 ^r dr' \; r'^2  \, C(r')  \ ,
\end{equation}
where we have used the simple relationship between density and nearest neighbours distance, $4\pi \rho = 3/r_1^3$.
In an infinitely large system, the bulk susceptibility is simply, $\chi_\infty = Q(r\to\infty)$, that is the full volume integral of the connected correlation function.
In a finite size system, however, due to the constraint, $\sum_i \vec{\delta \varphi_i} = 0$, we must have,
\begin{equation}
Q(r=L)=-1 \ ,
\end{equation}
for all systems, be they natural or synthetic, irrespective of the amount of real correlation. This relation is simply the mathematical consequence of the way velocity fluctuations are defined. Therefore, in a finite systems the susceptibility can be estimated as the maximum value reached by $Q(r)$ (this maximum is a lower bound for the bulk susceptibility). We know that, $C(r_0)=0$, so that the function $Q(r)$ reaches its maximum at $r=r_0$. Hence the finite size susceptibility is given by,
\begin{equation}
\chi = Q(r_0) = \frac{3}{r_1^3} \int_0^{r_0} dr \; r^2 \, C(r)  \ .
\end{equation}


\begin{table*}[b!]
\vskip 0.1 in
\begin{tabular}{l|c|c|c|c|c|c|c|c|c}
\hline
\hline
{\sc Species}   &\hspace{0.2cm}
 {\sc Event  label}   \hspace{0.2cm}     & \hspace{0.1cm}
 $N$    \hspace{0.2cm}     & \hspace{0.1cm}
 {\sc Duration (s)}    \hspace{0.2cm}     & \hspace{0.1cm}
 {l(mm)}    \hspace{0.2cm}     & \hspace{0.1cm}
 $r_1 (m)$    \hspace{0.2cm}     & \hspace{0.1cm}
 $r_0 (m)$    \hspace{0.2cm}     & \hspace{0.1cm}
 $|\vec{v}| (m/s)$    \hspace{0.2cm}     & \hspace{0.1cm}
 $\chi$    \hspace{0.2cm}     & \hspace{0.1cm}
 $\phi$
 \\
\hline
\hline
\multirow{3}{*}{\begin{minipage}[c]{4.8cm}\it Corynoneura scutellata \rm - CS \\ {\rm (Diptera:Chironomidae)} \\  \end{minipage}}
& 20110906\_A3 & 138 & 2.0  & 1.5 & 0.029 & 0.094 & 0.12 & 0.78 & 0.17 \\
& 20110908\_A1 & 119 & 4.4  & 1.1 & 0.036 & 0.105 & 0.13 & 0.46 & 0.27 \\
& 20110909\_A3 & 312 & 2.7  & 1.5 & 0.026 & 0.138 & 0.12 & 2.58 & 0.22 \\
\hline
\multirow{8}{*}{\begin{minipage}[c]{4.8cm}\it Cladotanytarsus  atridorsum \rm - CA \\ \rm  (Diptera:Chironomidae) \\  \end{minipage}}
& 20110930\_A1 & 173 & 5.9  & 2.4 & 0.057 & 0.228 & 0.23 & 1.48 & 0.31 \\
& 20110930\_A2 &  99 & 5.9  & 2.4 & 0.063 & 0.223 & 0.15 & 1.08 & 0.20 \\
& 20111011\_A1 & 131 & 5.9  & 2.4 & 0.075 & 0.272 & 0.11 & 0.65 & 0.17 \\
& 20120828\_A1 &  89 & 6.3  & 2.5 & 0.062 & 0.188 & 0.17 & 0.48 & 0.22 \\
& 20120907\_A1 & 169 & 3.2  & 1.9 & 0.062 & 0.330 & 0.13 & 1.72 & 0.20 \\
& 20120910\_A1 & 219 & 1.7  & 2.4 & 0.047 & 0.221 & 0.19 & 2.25 & 0.27 \\
& 20120917\_A1 &  192 & 0.36 & 2.2 & 0.043 & 0.219 & 0.12 & 2.09 & 0.14\\
& 20120917\_A3 & 607 & 4.23  & 2.2 & 0.033 & 0.259 & 0.10 & 5.57 & 0.15 \\
& 20120918\_A2 &  69 & 15.8 & 1.7 & 0.060 & 0.174 & 0.15 & 0.64 & 0.23 \\
& 20120918\_A3 & 214 & 0.89 & 1.7 & 0.041 & 0.230 & 0.20 & 2.04 & 0.36 \\
\hline
\multirow{7}{*}{\begin{minipage}[c]{4.8cm}\it Dasyhelea flavifrons \rm - DF \\ \rm (Diptera:Ceratopogonidae) \\  \end{minipage}}
& 20110511\_A2 & 279 & 0.9  & 2.3 & 0.053 & 0.248 & 0.20 & 1.25 & 0.35 \\
& 20120702\_A1 &  98 & 2.1  & 2.0 &0.062 & 0.162 & 0.14 & 0.69 & 0.20 \\
& 20120702\_A2 & 111 & 7.3  & 2.0 & 0.056 & 0.169 & 0.13 & 0.88 & 0.18 \\
& 20120702\_A3 &  80 & 10.0 & 2.0 & 0.060 & 0.170 & 0.12 & 0.32 & 0.20 \\
& 20120703\_A2 & 167 & 4.4  & 1.8 & 0.046 & 0.140 & 0.07 & 0.52 & 0.12 \\
& 20120704\_A1 & 152 & 10.0 & 1.7 & 0.050 & 0.154 & 0.09 & 0.63 & 0.15 \\
& 20120704\_A2 & 154 & 5.3  & 1.7 & 0.053 & 0.160 & 0.08 & 0.61 & 0.13 \\
& 20120705\_A1 & 188 & 5.9  & 1.8 & 0.055 & 0.182 & 0.12 & 0.92 & 0.20 \\
\hline
\hline
\end{tabular}
\caption{
{\bf Swarm data.} 
Each line represents a different swarming event (acquisition). $N$ is the number of individuals in the swarm, $r_1$ the time average nearest neighbour distance in the particular acquisition, $r_0$ the average correlation length, $\chi$ the average susceptibility and $\phi$ the average polarization.}
\label{tavolone}
\end{table*}
%
%
\clearpage

\clearpage

\section*{Legends for Supplementary Videos}

{\bf Video S1:}
Wild swarm of roughly $100$ midges in the field (Diptera:Ceratopogonidae). The swarm
has been video recorded at 170 frames per seconds, with a resolution
of 4Mpix, by a IDT-M5 camera.

\vspace{0.5 cm}
{\bf Video S2:}
Three dimensional visualization of the same natural swarm as in Video
S1. This 3d reconstruction has been obtained through the dynamical
tracking algorithm based on our trifocal experimental technique.

\vspace{0.5 cm}
{\bf Video S3:}
Three dimensional visualization of a numerically simulated swarm of
non-interacting particles in a harmonic potential (NHS-Noninteracting
Harmonic Swarm). The number of `midges' in the NHS is the same as in
Video S2.

\vspace{0.5 cm}
{\bf Video S4:}
Three dimensional visualization of a numerically simulated swarm obtained 
using Vicsek model with the addition of an harmonic attraction force towards the origin.
The video refers to a swarm in the ordered phase, with the polarization equal to $0.72$. 
The number of "midges" in the swarm is $128$, the harmonic constant $\beta$ is equal to $0.006$ while 
the simulation noise $\eta$ is equal to $0.3$.
\vspace{0.5 cm}

{\bf Video S5:}
Three dimensional visualization of a numerically simulated swarm obtained 
using Vicsek model with the addition of an harmonic attraction force towards the origin.
The video refers to a swarm in the disordered phase, with the polarization equal to $0.20$. 
The number of "midges" in the swarm is $128$, the harmonic constant $\beta$ is equal to $0.002$ while 
the simulation noise $\eta$ is equal to $0.45$.
\vspace{0.5 cm}

\end{document}